\documentclass[one,aps,final]{revtex4}
\usepackage{graphicx}

\begin{document}

\begin{abstract}
We report an experimental demonstration of effective entanglement in a
prepare\&measure type of quantum key distribution protocol. Coherent
polarization states and heterodyne measurement to characterize the transmitted
quantum states are used, thus enabling us to reconstruct directly their Q-function. By evaluating the excess noise of the states, we experimentally demonstrate that they fulfill a non-separability criterion previously presented by Rigas et al. [J. Rigas, O. G\"uhne, N. L\"utkenhaus, Phys. Rev. A {\bf 73}, 012341 (2006)]. For a restricted eavesdropping scenario we predict key rates using postselection of the heterodyne measurement results.
\end{abstract}

\title{Witnessing effective entanglement in a continuous variable prepare\&measure setup and application to a QKD scheme using postselection}

\author{S.~Lorenz$^{*}$, J.~Rigas$^{\dagger}$, M.~Heid$^{\dagger}$, U.L.~Andersen$^{*}$, N.~L\"utkenhaus$^{\dagger}$, G.~Leuchs$^{*}$}

\address{$^{*}$Institute of Optics, Information and Photonics,
Universit\"at Erlangen-N\"urnberg,
G\"unther-Scharowsky-Stra{\ss}e 1/Bau 24, D-91058 Erlangen, Germany. \\
$^{\dagger}$Institute of Theoretical Physics I,
Universit\"at Erlangen-N\"urnberg,
Staudtstra{\ss}e 7/B2, D-91058 Erlangen, Germany.}
\date{\today}

\pacs{03.67.Mn, 42.50.Lc, 42.25.Ja, 03.67.Dd}

\maketitle

\section{Introduction}
The process of Quantum Key Distribution (QKD) \cite{WIE83,BB84} uses quantum
mechanical properties of light fields to establish a secret shared key between
two honest parties, named Alice and Bob. This key is then used to ensure
secret communication between Alice and Bob by means of a classical cipher like
the one-time-pad \cite{Vernam}. The adversary of Alice and Bob is an
eavesdropper Eve, who tries to gain the maximum knowledge about the key
without being noticed by Alice and Bob. Eve can use any method within the laws
of quantum mechanics, and therefore is not restricted by technological imperfections. 

The physical implementation of a QKD protocol requires two channels between Alice and Bob. Over the \textit{quantum channel} Alice and Bob can exchange quantum states. By the laws of quantum mechanics Alice and Bob are able to detect any interference of Eve with the quantum states. Classical information is exchanged on the \textit{classical channel}. This channel has to be authenticated in order to prevent a man-in-the-middle attack by Eve.

After the quantum states have been exchanged over the quantum channel, they
are measured by Bob. Alice and Bob keep the results of the preparation process and the measurement process,
thus sharing a set of classical correlated measurement data described by the
joint probability distribution $p(A;B)$. This is the first stage of the QKD
protocol. In the second stage, Alice and Bob try to generate a key pair from
their correlations $p(A;B)$ and correct possible errors. From the disturbance of the correlations they deduce the amount of information Eve might have on the key pair, and reduce Eve's information by privacy
amplification. For these tasks, only communication over the classical channel is
needed, as all exchanged information is classical. If the QKD is successful, Alice and Bob will share a key and have an upper bound on the information Eve might have about it.

It has been shown by Curty et al. \cite{CUR04b, CUR05} that there is a
necessary precondition for the second stage to succeed: The correlations
$p(A;B)$ coming from the first stage have to be created from an effective
entangled quantum state shared between Alice and Bob. Only then it is possible
to generate a secret key from the data set. Note that this 'effective
entanglement' does not mean that entanglement as a physical resource has to be
used in the state preparation step. It is sufficient that Alice and Bob can model their correlations as if they had shared an entangled state. We use an entanglement witness to check if the correlations show effective entanglement.

In this paper, we demonstrate this effective entanglement for a particular
implementation of the quantum channel. We present a prepare\&measure type
setup, which uses the polarization of coherent light pulses to generate
nonorthogonal quantum states. The pulses are characterized by a heterodyne
measurement \cite{SHA84} on Bob's side, allowing for a reconstruction of their
antinormal ordered quasi-distribution, or Q-function \cite{WAL87}. We show for
this particular system that we can prove effective entanglement for the
prepared quantum states using a model developed by Rigas et
al. \cite{Rigas,RIG05}. Thus, it has clearly the potential of generating secret
keys. Following this, we use a reasonable model to predict key rates
for a classical key generation process with the data obtained in stage one of
the experiment. It considers postselection \cite{SIL02b} with direct
or reverse reconciliation \cite{GRO02b} of the measurement data. This QKD protocol is known
to be secure against an adversary Eve who is restricted to beam splitting
attacks \cite{HEI05}.

The paper is divided into five subsections. In section II we briefly introduce the theoretical background of the entanglement verification process, as it is described by Rigas et al. \cite{Rigas, RIG05}. 
We also present the QKD quantum state protocol there. In section III we give a characterization of the experimental apparatus which implements the quantum stage of the QKD system. Section IV shows how the Q-function can be experimentally reconstructed and how  the effective entanglement of the QKD setup can be verified. Section V gives achievable key rates for our experiment applying a postselection procedure.


\section{Prepare\&measure quantum key distribution and verification of effective entanglement}
The existing QKD systems fall into two categories: entanglement based systems and prepare\&measure systems. For a review on both see e.g. \cite{GIS02}. In entanglement based systems, a bipartite entangled state is produced by a source which might even be under Eve's control. One part of the entangled state is then sent to Alice while the other is sent to Bob. Here Alice and Bob can directly verify the entanglement of the state, thus bounding any interaction of Eve \cite{EKE91}. Then privacy amplification can be understood as an entanglement distillation (see e.g. Shor and Preskill \cite{SHO00}). 

In a prepare\&measure system Alice prepares a quantum state and sends it through the quantum channel to Bob \cite{BB84, BEN92}. He characterizes the quantum state, and from Alice's preparation and Bob's measurement results they estimate Eve's action and information on the quantum state. As sources of entangled states are hard to implement and suffer from technical disadvantages, we use the latter approach in our experiment, which ensures stable, deterministic state preparation with minimum resources.

In our protocol Alice encodes her bit values into two nonorthogonal states as in the B92 protocol \cite{BEN92}. In particular, she prepares coherent states with the amplitudes $-\alpha$ or $+\alpha$. A general coherent state can be described in a Fock state basis by
\begin{equation}
	|\alpha \rangle = e^{-\frac{|\alpha|^2}{2}} \sum_n \frac{\alpha^n}{\sqrt{n!}}|n\rangle
\end{equation}
where $n$ is the photon number, and $|n\rangle$ a photon number
eigenstate. The coherent states constitute an overcomplete basis set, because there
is always an overlap between two coherent states with amplitudes $\alpha$ and
$\beta$, as given by
\begin{equation}
	\langle \beta | \alpha \rangle = e^{-\frac{1}{2}| \beta - \alpha |^2}.
	\label{Eqn_Overlap}
\end{equation}
Thus it is impossible to discriminate between the coherent states $|-\alpha\rangle$ and $|+\alpha\rangle$ with certainty \cite{FUC96,HUT96,IVA87,FUC00}. The coherent states emitted by Alice are transmitted through the quantum channel, which is under the control of the eavesdropper Eve. She can manipulate the states and adjust the channel properties to get an advantageous position in the key generation process. As the states enter Bob's measurement station, he performs a heterodyne measurement \cite{SHA84}, in contrast to the original B92 setup, where a photon counter is used to discriminate between the different states. The heterodyne measurement splits the optical mode on a 50/50 beam splitter and measures the two conjugate field quadratures on its outputs with two homodyne detectors. This measurement of two conjugate observables (see e.g. \cite{ART65,STE92}) corresponds to a projection on coherent states. The quadrature operators are derived from the creation and annihilation operators $a^\dagger$ and $a$ by:
\begin{equation}
	\mathbf{X}=\frac{1}{2}\left( a^\dagger + a \right)
	;\quad \mathbf{Y}=\frac{i}{2}\left( a^\dagger - a \right).
\end{equation}
Bob records the results of the heterodyne measurement in a two-dimensional histogram, which represents the Q-function \cite{HUS40,WAL87,VOG89,LAI89,LEO93a,LEO93b}:
\begin{equation}
	Q(\mathrm{Re} \beta; \mathrm{Im} \beta)=\frac{1}{\pi} \langle \beta | \rho | \beta \rangle.
	\label{Qfunction}
\end{equation}
Here, the general state $\rho$ is projected onto the coherent state $|\beta\rangle$.

\begin{figure}[thb]
\begin{center}
\includegraphics[width=0.35\columnwidth]{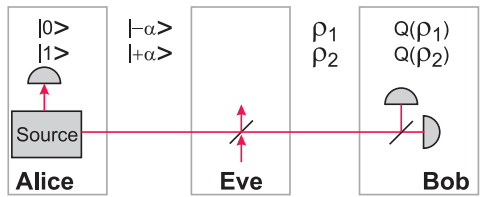}
\end{center}
\caption{Simplified theoretical setup.} 
\label{Figure_Infotheorie_Setup}
\end{figure}

To model the prepare\&measure setup with effective entangled states, we follow
Bennett et. al. \cite{BEN92b} and assume that Alice possesses a source of
bipartite quantum states given by
\begin{equation}
	|\Psi\rangle_{\mathit{Alice\rightarrow Bob}}=\frac{1}{\sqrt{2}}|\Psi_0\rangle_{AB}+\frac{1}{\sqrt{2}}|\Psi_1\rangle_{AB},
\end{equation} 
whereas the states $|\Psi_{0,1}\rangle_{AB}$ are given by
\begin{eqnarray}
	|\Psi_0\rangle_{AB}&=&|0\rangle_\mathrm{Alice}\otimes|-\alpha\rangle_\mathrm{to~Bob}\;, \nonumber \\ 
	|\Psi_1\rangle_{AB}&=&|1\rangle_\mathrm{Alice}\otimes|+\alpha\rangle_\mathrm{to~Bob}\;.
\end{eqnarray}
The model setup is shown in Fig.~\ref{Figure_Infotheorie_Setup}. Alice keeps
the first part of the state, which consists of a qubit, and sends the other
part of the state over the quantum channel to Bob. By detecting her qubit,
Alice effectively prepares a coherent state of amplitude $-\alpha$ or
$+\alpha$ as signal.
Thus, conditioned on her qubit measurement result, Alice produces a certain coherent state. As her measurement result, 0 or 1, is occurring completely randomly, but also completely correlated with the generated coherent state, the entanglement source resembles the random production of coherent states with amplitudes $-\alpha$ or $+\alpha$. In reality, such a random production of coherent states can be enabled without the use of an entanglement source, but in the theoretical description there is no difference between these two physical systems.

The coherent states travel over the quantum channel, where Eve can interact with them. Also channel losses are attributed to Eve, as she can always replace a lossy channel with a lossless one and tap off the surplus intensity. When Eve has interacted, the pure coherent states might have changed to more general mixed states described by the density matrices $\rho_1$ and $\rho_2$. From the results of the heterodyne measurement, Bob can reconstruct the Q-function, and deduce the quadrature variances $\Delta^2 \mathbf{X}=\langle \mathbf{X}^2 \rangle - \langle \mathbf{X} \rangle^2$ and $\Delta^2 \mathbf{Y}=\langle \mathbf{Y}^2 \rangle - \langle \mathbf{Y} \rangle^2$ and all other elements of the covariance matrix.

As the full joint density matrix of Alice and Bob is not accessible by heterodyne measurements and dichotomic preparation, we revert to the bipartite expectation value matrix defined in \cite{RIG05}
to describe the state shared by Alice and Bob. It consists of a part A describing Alice's state preparation, and a part $\mathbf{B}$ describing Bob's heterodyne measurement
\begin{equation}
\chi = \left[
\begin{array}{cc}
\langle \; |0\rangle \langle 0|_\mathrm{A} \otimes \mathbf{B}  \; \rangle & \langle \; |1\rangle \langle 0|_\mathrm{A} \otimes \mathbf{B}  \; \rangle \\
\langle \; |0\rangle \langle 1|_\mathrm{A} \otimes \mathbf{B}  \; \rangle & \langle \; |1\rangle \langle 1|_\mathrm{A} \otimes \mathbf{B}  \; \rangle\\
\end{array} \right],
\label{Eqn_chi-Matrix}
\end{equation}
with the matrix $\mathbf{B}$ composed of the quadrature operators directly accessible to Bob:
\begin{equation}
\mathbf{B}=\left[
\begin{array}{ccc}
1 & X & Y \\
X & X^2 & \frac{1}{2}(XY+YX) \\
Y & \frac{1}{2}(XY+YX) & Y^2 \\
\end{array} \right].
\label{Eqn_B-Matrix}
\end{equation}

\begin{figure}
\begin{center}
\includegraphics[width=0.4\columnwidth]{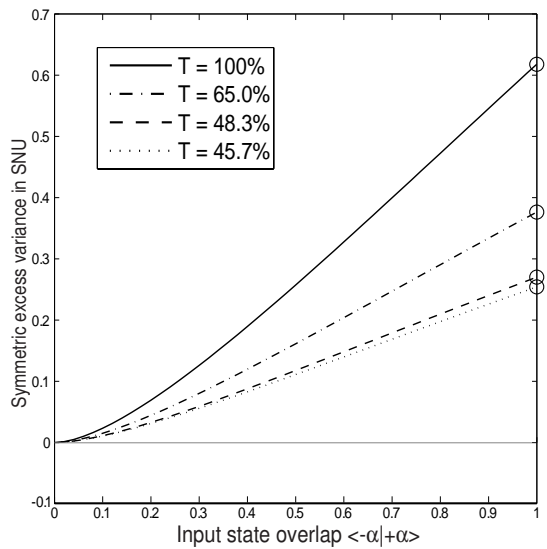}
\end{center}
\caption{Graphical representation of the entanglement criterion. For excess variances $E$ (in shot noise units: SNU) below the curves, the correlated data $p(A;B)$ cannot be explained by separable states. Zero excess variance corresponds to the detection of a pure coherent state. Different curves belong to different quantum channel transmissions $T$.} 
\label{Figure_Ent_Criterion_Theory}
\end{figure}

It has been shown \cite{RIG05}, that an expectation value matrix with certain restrictions
can only be justified by assuming that an entangled state is shared between Alice and Bob. This
is done by using a separability criterion (positive partial transposition
type). To verify this, only the blocks on
the diagonal of $\chi$ have to be computed. These are given by the expectation values corresponding to matrix $\mathbf{B}$ for the two conditional signal states and are therefore completely characterized by Bob's quadrature measurements. By proving this effective entanglement from the
experimental data \cite{RIG05}, we can fulfill the first precondition to
generate a secret key from Bob's and Alice's correlations. The separability
condition can be evaluated by semi-definite programming \cite{VAN96}, giving an upper bound
on the tolerable noise $\Delta^2 \mathbf{X}, \Delta^2 \mathbf{Y}$ below which the effective entanglement can be
verified. This bound depends on the input state overlap $\langle - \alpha | +
\alpha \rangle$ and the quantum channel loss. We define the excess noise or
excess variance $E$ of an observable $\mathbf{X}$ for a signal state by
comparing its variance to the variance of a coherent vacuum state (shot noise) as
\begin{equation}
	E(\mathbf{X})=\frac{\Delta^2 \mathbf{X}(\mathrm{signal})}{\Delta^2 \mathbf{X}(\mbox{vacuum})}-1.
	\label{Eqn_ENV}
\end{equation}
Figure \ref{Figure_Ent_Criterion_Theory} shows the numerically calculated bounds to the excess variance for different quantum channel transmissions.
All experimental excess variances which are below their corresponding bounds fulfill the non-separability condition and thus the scheme exhibits effective entanglement.


\section{Experimental apparatus}
The experimental setup deviates from the theoretical description in the
previous section in one aspect. To determine the quadratures $\mathbf{X}$ and
$\mathbf{Y}$, Bob has to use a phase reference as local oscillator for the
homodyne measurements \cite{YUE80, SHA84}. This phase reference is sent along
with the signal state in our experiment, as it is done in most continuous
variable quantum cryptography experiments \cite{HIR00, COR03, GRO03, LOR04, WEE04, LAN05}. The analysis of quantum correlations does not directly apply
to this new situation. However, there is no obvious possibility for Eve to manipulate the local oscillator in her favor, as it is
a classical signal, and its phase can be publicly announced by Alice and Bob
without giving away any information about the signal state. Moreover, Bob can
use his own local oscillator and phase lock it to Alice's local oscillator, as
described in \cite{KOA04}, so that all that Eve can do to the local oscillator
is to introduce phase jitter, which disturbs Bob's measurements without giving
Eve any information on Alice's quantum states. In that case, the previous
analysis would apply again. From now on we assume that the local oscillator is not manipulated by Eve in any way.

Consequently our experimental realization of the quantum channel has to transmit two light modes: the signal field mode $\mathbf{a}$ contains the weak coherent pulses in which the quantum information is encoded. The local oscillator mode $\mathbf{b}$ is needed in the heterodyne measurement of the signal mode as a phase reference. Instead of using two spatially separated channels to transmit the two modes, we use two orthogonal polarization modes as representing the two fields in one spatial mode. This facilitates the generation of the signal states as well as providing high quality interference of the two modes at the heterodyne detector. The amplitude and relative phase of the two orthogonal polarization modes can be described by the Stokes parameters \cite{STO52} or by the Stokes operators \cite{FAN49,COL70,KOR02} in quantum theory. In our notation they read:
\begin{eqnarray}
	\mathbf{S_0} &=& a^\dagger a + b^\dagger b , \\
	\mathbf{S_1} &=& a^\dagger a - b^\dagger b , \\
	\mathbf{S_2} &=& a^\dagger b + b^\dagger a , \\
	\mathbf{S_3} &=& -i(a^\dagger b - b^\dagger a ).
\end{eqnarray}
The intensity in the local oscillator mode $\mathbf{b}$ is always much larger than the intensity in the signal mode $\mathbf{a}$, thus we have $\langle S_0 \rangle \simeq - \langle S_1 \rangle \gg 0, \langle S_2 \rangle  \approx 0, \langle S_3 \rangle \approx 0$. The relative phases and amplitudes of the polarization modes can be manipulated with birefringent optical elements and polarizing beam splitters. 

\begin{figure}
\begin{center}
\includegraphics[width=0.35\columnwidth]{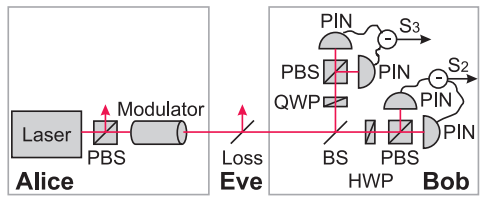}
\end{center}
\caption{Simplified experimental setup. Alice prepares coherent polarization states with a cw laser diode and a Faraday modulator. Bob characterizes the incoming light by a heterodyne measurement consisting of two polarization sensitive homodyne setups. Eve is simulated by changing the loss of the quantum channel.} 
\label{Figure_Setup}
\end{figure}

A schematic drawing of our setup is shown in Fig.~\ref{Figure_Setup}.
An external cavity laser diode emits continuous wave light at 810nm and by a
polarizing beam splitter acting as polarization filter a coherent bright state
is created in the local oscillator mode $\mathbf{b}$ whereas the signal mode
$\mathbf{a}$ is in the vacuum state. The light passes through an magneto
optical modulator that utilizes the Faraday effect to alter the polarization
state \cite{Yariv}. Depending on the externally applied magnetic field, the
modulator rotates a linearly polarized input field and shifts the phase of
circular polarized fields. The light polarization can be varied continuously
from $\mathbf{S_1}$-polarized to $\mathbf{S_2}$-polarized, which corresponds
to equal optical power in the $\mathbf{a}$ and the $\mathbf{b}$ modes. In our
case, we only induce a very tiny modulation, such that for the optical powers
$P_a, P_b$ in the two modes $P_b \gg P_a$ is always satisfied. The modulation
is applied in pulses of 5$\mu$s duration, either with parallel or antiparallel
magnetic field orientation, such that either the state $|+\alpha\rangle$ or
the state $|-\alpha\rangle$ is produced in the $\mathbf{a}$ mode. The state
overlap $\langle +\alpha|-\alpha \rangle = e^{-2|  \alpha |^2}$ is in the
range from 0.2 to 0.8. When encoding the signal, the intensity in the local
oscillator mode is reduced only by a negligible amount (intensity variations
are smaller than $10^{-9}$ in our experiment). Therefore, a local oscillator of constant power can be assumed. As the signal field is derived from the local oscillator field through modulation, the relative phase of both fields is constant, even though the laser phase might suffer from fluctuations.

The beam is then directed to Bob, traveling through Eve's domain over a free
space link of approximately 20cm. Various quantum channel transmissions can be simulated by using neutral density filters to equally attenuate both modes $\mathbf{a}$ and $\mathbf{b}$.

\begin{figure}
\begin{center}
\includegraphics[width=0.35\columnwidth]{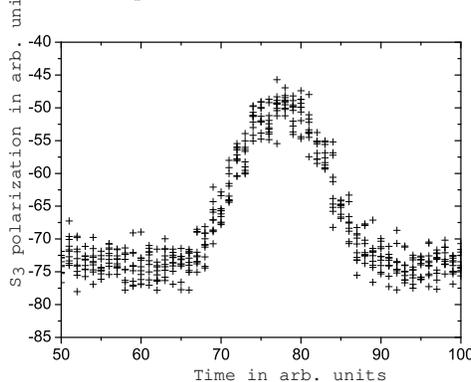}
\end{center}
\caption{Demonstration of a detected signal in the time domain. Ten identical measurements were superimposed to show the variations in detector voltage due to quantum noise. The pulse duration was set to 50$\mu$s at a sample rate of 1MSample/s to better visualize the pulse shape and the discrete nature of the sampled data. In the further experiments, a pulse duration of 5$\mu$s was used.} 
\label{Figure_Demo_Trace}
\end{figure}

In Bob's receiver, the incoming beam undergoes a heterodyne measurement. It is split on a polarization independent 50/50 beam splitter, and both parts are directed to individual homodyne measurement setups, which record the $\mathbf{S_2}$-polarization and $\mathbf{S_3}$-polarization respectively. This is done by interfering the signal and the local oscillator modes on a beam-splitter and subsequently recording the intensity difference at the beamsplitter output ports \cite{FOR61, MAN66, YUE80}. As long as the local oscillator mode is much brighter than the signal mode, the difference photocurrent $I$ corresponds to 
\begin{equation}
	I \propto \sqrt{P_b} \sqrt{P_a} \cos \phi
	\label{Eq_Homodyn}
\end{equation}
with $P_b$ being the local oscillator optical power and $P_a$ the signal optical power, and $\phi$ the relative phase between signal and local oscillator. The relative phase of signal and local oscillator in our setup is controlled by the appropriate choice and setting of half wave plates (HWP, $\mathbf{S_2}$ measurement) and quarter wave plates (QWP, $\mathbf{S_3}$ measurement). The two modes interfere at polarizing beam splitters. As both the signal and the local oscillator are in the same spatial mode, a very high interference contrast can be achieved. We record a polarization contrast larger than $10^4$, corresponding to an interference visibility larger than 99.9\%. Both homodyne detector voltages are sampled with a fast A/D converter. Figure \ref{Figure_Demo_Trace} shows 10 superimposed test pulses with high amplitude in the $\mathbf{a}$ mode. Each point corresponds to one sample. It can be seen that one polarization pulse is much longer than one sample period. The variance of the sampled data is an indicator for the shot noise of the signal light mode.
For the experiments, a sample rate of 20MSample/s and a pulse duration of 5$\mu$s was chosen. Consequently, an integration over 100 Samples defines our pulse amplitude. The electronic and dark noise of the detectors is more than 14dB below the signal noise in a frequency window from 100Hz to 2MHz with a local oscillator power of $P_b$=1.2mW. Dark noise at higher frequency is filtered by a 2MHz low pass filter. From the pulse amplitudes, the quadratures and polarizations are calculated by measuring the local oscillator power, and the detector transimpedance (approx. 110k$\Omega$) as well as the diode quantum efficiencies (91\% $\pm$ 3.5\%). 

With the pulse separation of 10$\mu$s a clock rate of 100kPulse/s is feasible. As we characterized five vacuum noise time slots for each bright signal pulse, the effective clock rate for signal pulses was reduced to 16.7kPulse/s for the Q-function and effective entanglement measurements. In the real QKD system, the vacuum characterization is only needed for an initial calibration, and the full clock rate can be used for pulse transmission subsequently.

A whole measurement sequence consists of 250000 signal pulses. 
\begin{figure}
\begin{center}
\includegraphics[width=0.35\columnwidth]{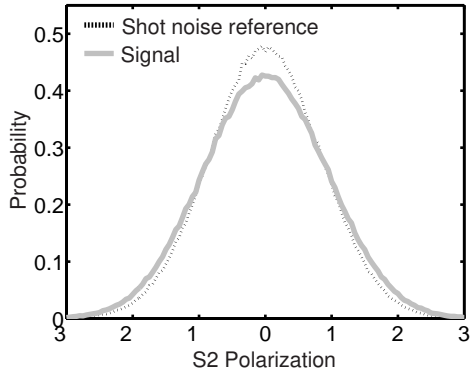}
\end{center}
\caption{Marginal distribution of the measured polarization values. The dotted black curve represents the shot noise reference, recorded with vacuum in the signal mode $a_S$. The solid grey curve is Bob's measurement value histogram for two alternating states, corresponding to Fig.~\ref{Figure_MixedStateBob} and the highlighted line in Table~\ref{Table_Variances}.}
\label{Figure_Marginals}
\end{figure}
A histogram of recorded pulse amplitudes in the $\mathbf{S_2}$ polarization (corresponding to a 0 or $\pi$ phase shift between signal mode $\mathbf{a}$ and local oscillator mode $\mathbf{b}$) is shown in Fig. \ref{Figure_Marginals}. The shot noise reference is produced with no modulation current (vacuum mode) and is shown in black (dotted). The grey histogram is derived from the signal pulses with amplitudes $-\alpha$ and $+\alpha$. The overlap of the two resulting Gaussian distributions is too high to distinguish between them in this histogram, the only visible effect is a decrease in peak height and an increase in variance.

The variances of the polarization (or quadrature of the $\mathbf{a}$ mode)
measurement can be used to calculate the appropriate entries of the
$\chi$-matrix (cf. Eqn.~\ref{Eqn_chi-Matrix}). The Q-functions were
reconstructed by building a histogram of the $\mathbf{S_2}$ and $\mathbf{S_3}$
values and renormalizing the volume of the resulting two dimensional
function. From the definition of the Q-function of Eqn.~\ref{Qfunction} it
follows that its peak value is maximal for pure coherent states. Thus a rough measure for the purity of the states measured by Bob is the peak height of his Q-function for each type of state Alice prepared. The absolute height of the measured functions is in accordance with the total losses measured for diode efficiency and optical losses in the detection setup, which sum up to 13.6\%. It is assumed that these losses cannot be actively used by Eve. For the entanglement criterion, the excess noise variance (cf. Eqn.~\ref{Eqn_ENV}) is used, which compares the signal variance with the vacuum variance. As the vacuum variance is determined with the same setup, the detection losses are not regarded in the further analysis.


\section{Results}
\begin{figure}
\begin{center}
\includegraphics[width=0.4\columnwidth]{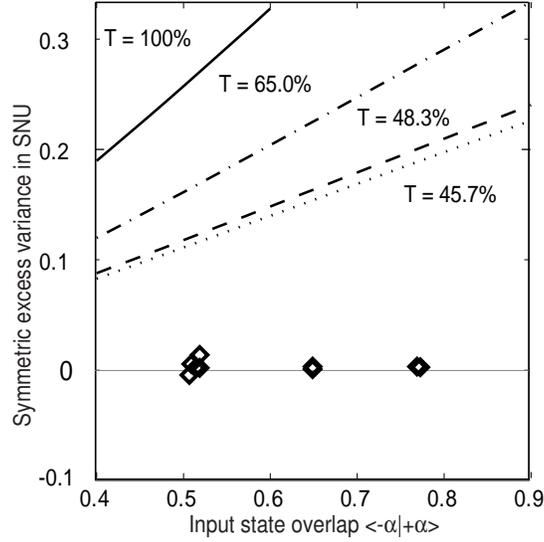}
\end{center}
\caption{Graphical representation of the entanglement criterion. For excess variances below the curves, the correlated data $p(A;B)$ cannot be satisfied by separable states. Different curves correspond to different quantum channel losses. The open diamonds show measured averages of $E(\mathbf{S_2})$ and $E(\mathbf{S_3})$, their numerical values can be seen in Table~\ref{Table_Variances}.}
\label{Figure_Ent_Criterion}
\end{figure}

The restrictions for the quadrature variances given by the entanglement criterion of Rigas et al \cite{RIG05} are shown in Fig.~\ref{Figure_Ent_Criterion} for the relevant parameter range.
The curve shows the maximum measured excess variance $E$ compared to the coherent vacuum state, that is tolerable without having a separable state. As it can be seen, all measurement results (diamonds) lie below this threshold. Therefore the joint probability distribution can only be explained by effective entanglement in the whole shared state between Alice and Bob. Numerical values are compiled in Table~\ref{Table_Variances}.
\begin{table}
\begin{tabular}{ccccc}
State & Quantum  & Excess  & Excess & $E(\mathbf{S_2})$\\ 
overlap & channel & variance & variance & with electronic\\
$\langle -\alpha | + \alpha \rangle$ & transmission T & $E(\mathbf{S_2})$ & $E(\mathbf{S_3})$ & noise subtracted\\ \hline \hline
0.51 & 100\% & 0.8\% & 0.1\% & 4.5\%\\ 
\textbf{0.50} & \textbf{45.7\%} & \textbf{0.2\%} & \textbf{-0.3\%} & \textbf{8.3\%}\\ 
\hline
0.78 & 100\% & -0.2\% & -0.2\% & 3.5\% \\ 
0.77 & 45.7\% & 0.3\% & 0.1\% & 8.3\% \\ 
\hline
0.52 & 100\% & 1.6\% & 0.1\% & 5.3\% \\ 
0.52 & 48.3\% & 0.2\% & 0.3\% & 7.7\% \\ 
0.51 & 65.0\% & 0.2\% & 0.0\% & 5.8\% \\ 
\hline
0.65 & 100\% & 0.6\% & -0.5\% & 4.2\% \\ 
0.65 & 48.3\% & 0.1\% & 0.0\% & 7.5\% \\ 
\hline \hline
\end{tabular}
\caption{Excess variances $E$ for the coherent state measurement, depending on state overlap and quantum channel transmission. The statistical error is $\pm0.5$ for $E(\mathbf{S_2})$ and $\pm0.3$ for $E(\mathbf{S_3})$. The last column gives the excess variance with the electronic noise subtracted only from the shot noise reference which refers to a worst case scenario. The highlighted line shows the data set that produced the Q-function in Fig.~\ref{Figure_MixedStateBob} and the marginal distribution in Fig. \ref{Figure_Marginals}.}
\label{Table_Variances}
\end{table}
For each measurement, a separate evaluation of the vacuum variance (shot noise
level) was calculated from the vacuum pulses transmitted during the
measurement. The state overlap prepared by Alice is shown in the first
column. The second column gives the quantum channel transmission, where losses
in Bob's detection unit are not taken into account. The third column gives the
excess variance $E(\mathbf{S_2})$ of Bob's polarization measurements compared
to the vacuum variance. The fourth column gives the excess variance
$E(\mathbf{S_3})$ of the $\mathbf{S_3}$ polarization. Apart from one value,
all excess variances fall well below 1\%, whereas more than 10\% are enough to
prove effective entanglement with the given quantum channel
transmissions. Note that negative excess variances are no sign of nonclassical
states but represent the statistical variations due to the finite sample
size. The average excess variance is given by $E(\mathbf{S_2})=0.4\pm0.5\%$
and $E(\mathbf{S_3})=0.0\pm0.3\%$. To give a conservative estimate on the
impact of electronic noise on our measurement results, we subtracted the
variance of the electronic noise \textit{only} from the shot noise reference. The excess variances are compared to this corrected vacuum state are given in column five. They are all still below 9\%. Even with this conservative correction we witness the presence of effective entanglement.

\begin{figure}
\begin{center}
\includegraphics[width=0.4\columnwidth]{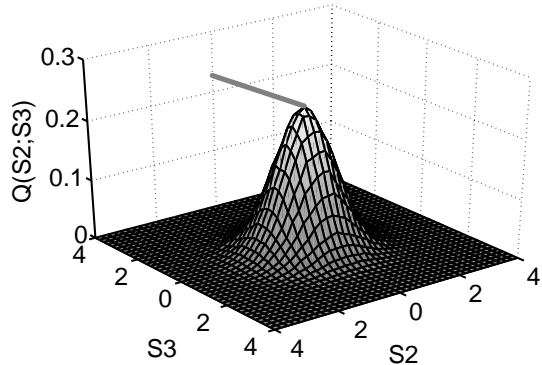}
\end{center}
\caption{Vacuum noise Q-function. $\mathbf{S_2}$ and $\mathbf{S_3}$ are proportional to the $\mathbf{X}$ and $\mathbf{Y}$ quadrature of the signal mode $a$. The peak height is indicated by the grey line.}
\label{Figure_VacuumNoise}
\end{figure}

\begin{figure}
\begin{center}
\includegraphics[width=0.4\columnwidth]{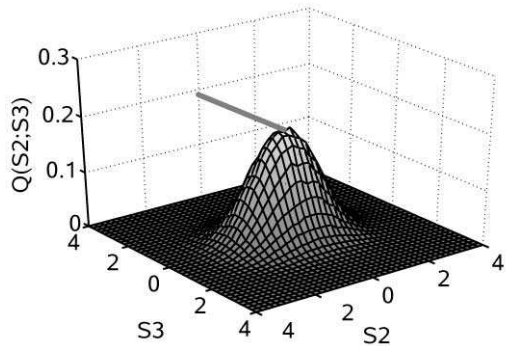}
\end{center}
\caption{Mixed Q-function of the two signal states $\rho_1$ and $\rho_2$ after leaving Alice's preparation.} 
\label{Figure_MixedStateAlice}
\end{figure}

\begin{figure}
\begin{center}
\includegraphics[width=0.4\columnwidth]{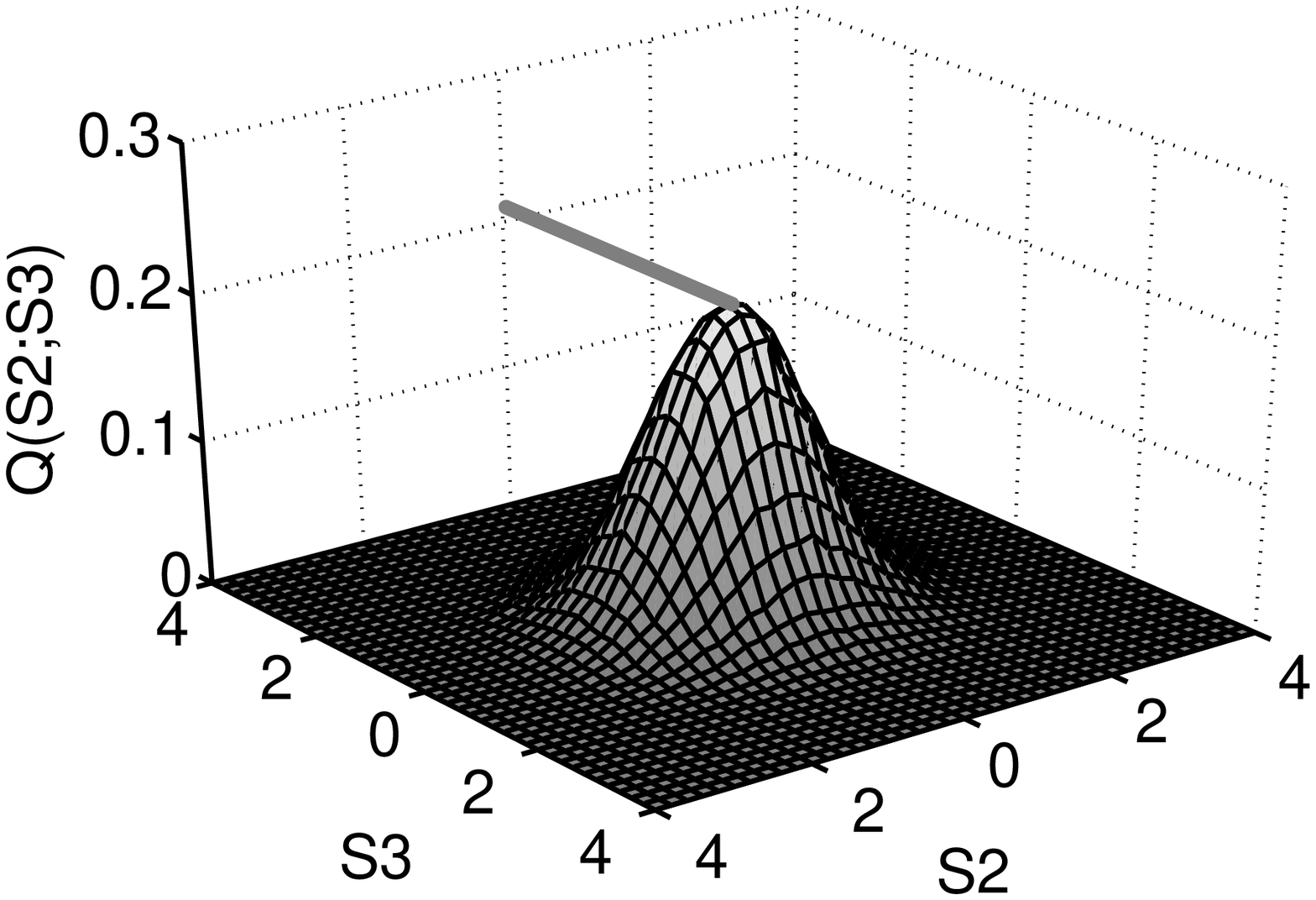}
\end{center}
\caption{Mixed Q-function of the two signal states $\rho_1$ and $\rho_2$ in balanced mixture after experiencing 54.3\% of channel loss. This Q-function corresponds to the highlighted line in Table~\ref{Table_Variances}.} 
\label{Figure_MixedStateBob}
\end{figure}

Fig.~\ref{Figure_VacuumNoise} shows the reconstructed Q-function of the vacuum state.
This function is a direct histogram, and has not been smoothed or fitted. In Fig.~\ref{Figure_MixedStateAlice} we depict the Q-function of the mixed state $\rho=\frac{1}{2}|+\alpha\rangle\langle +\alpha| + \frac{1}{2}|-\alpha\rangle\langle -\alpha| $ with an overlap $\langle -\alpha|+\alpha \rangle$=0.51 measured with no channel loss. From the figure it can be seen that the height of the Q-function is an indicator to mixedness of the depicted state. Here the mixture of $\rho_1$ and $\rho_2$ states with no additional quantum channel loss gives a Q-function with a peak height that is distinctively less than that of the pure vacuum state.
After a loss of 54.3\%, the mixedness of the state is decreased (cf. Fig.~\ref{Figure_MixedStateBob}). This Q-function corresponds to the highlighted line in Table~\ref{Table_Variances} and the grey histogram in Fig.~\ref{Figure_Marginals}.

\begin{figure}
\begin{center}
\includegraphics[width=0.4\columnwidth]{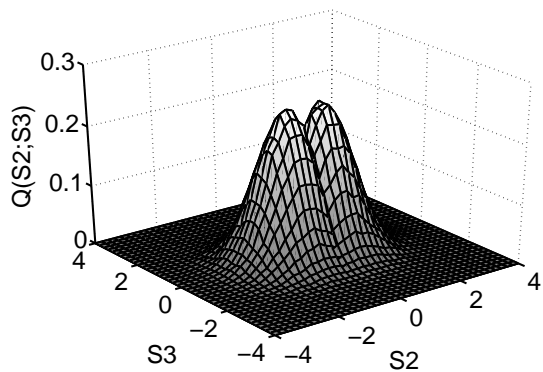}
\end{center}
\caption{Both Q-functions of the $\rho_1$ and the $\rho_2$ state, measured directly after preparation. This figure corresponds to Fig.~\ref{Figure_MixedStateAlice}.} 
\label{Figure_TwoStateAlice}
\end{figure}

\begin{figure}
\begin{center}
\includegraphics[width=0.4\columnwidth]{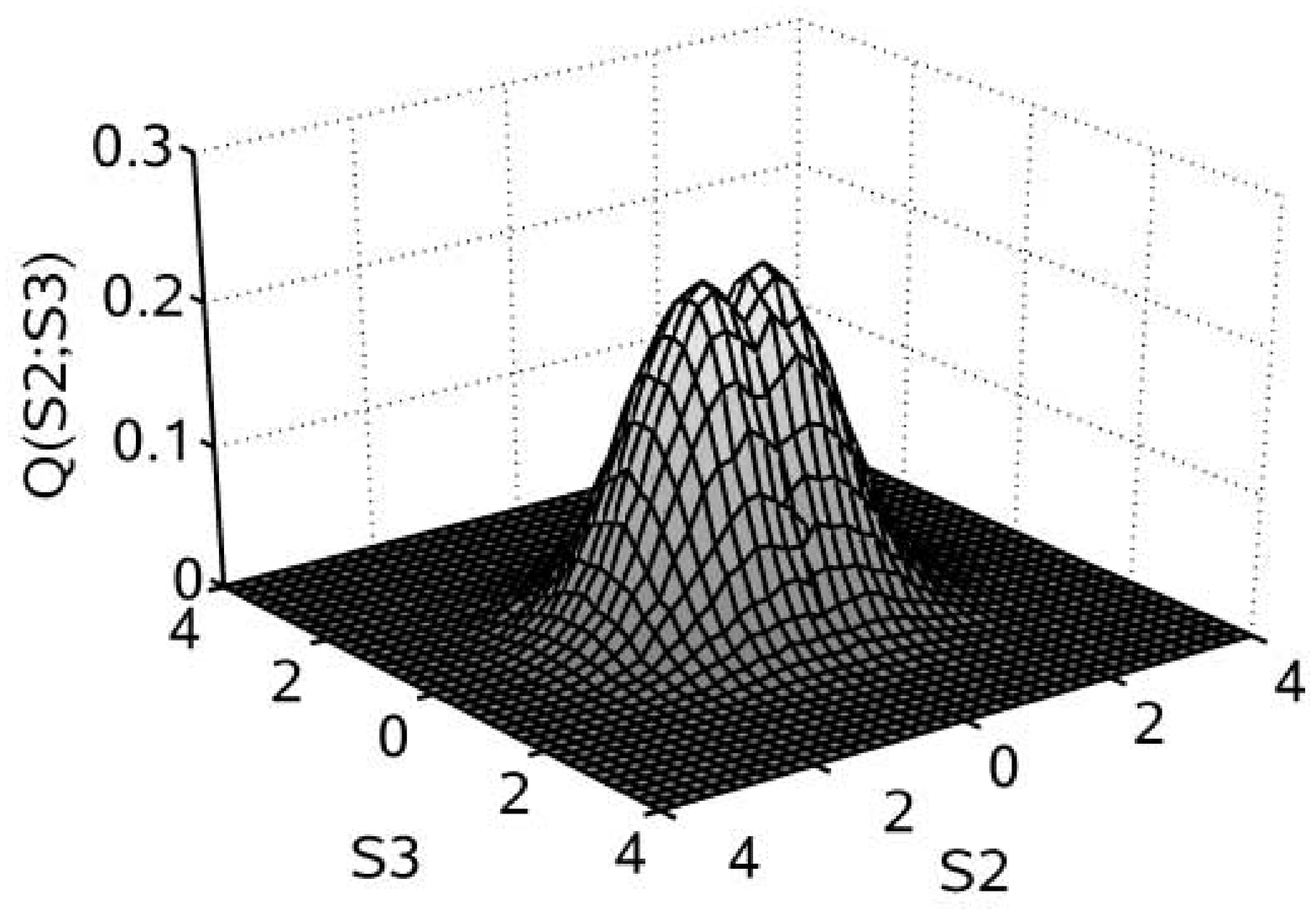}
\end{center}
\caption{Both Q-functions of the $\rho_1$ and the $\rho_2$ state, measured after propagation with 54.3\% losses. This figure corresponds to Fig.~\ref{Figure_MixedStateBob}.} 
\label{Figure_TwoStateBob}
\end{figure}

If Alice reveals to Bob which pulse belonged to the state
$|\Psi_0\rangle_{AB}$ and which to the state $|\Psi_1\rangle_{AB}$, Bob can produce two histograms. They are both shown in Fig.~\ref{Figure_TwoStateAlice}. 
The two Gaussian distributions represent the states Bob receives of $|\Psi_0\rangle_{AB}$ ($\rho_1=|-\alpha\rangle\langle -\alpha|$ ideally) and of $|\Psi_0\rangle_{AB}$ ($\rho_2=|+\alpha\rangle\langle+\alpha|$ ideally). The overlap (cf. Eqn.~\ref{Eqn_Overlap}) can be seen at the intersection of the two distributions ($\mathbf{S2}$ coordinate is zero). This overlap increases as losses are introduced. Fig.~\ref{Figure_TwoStateBob} shows the two states' Q-functions after 54.3\% losses. 
The two Gaussians moved closer together, the overlap has increased. In the ultimate limit of 100\% loss, a pure vacuum state (cf. Fig.~\ref{Figure_VacuumNoise}) would be registered by Bob.


\section{Application to a continuous variable QKD scheme}
We now apply our prepare\&measure system to a specific quantum key
distribution scheme. Alice prepares either the coherent state
$|+\alpha\rangle$ or $|-\alpha\rangle$ as signal. As shown in the previous
sections, Alice and Bob can verify the entanglement in their virtually shared
state by simultaneously measuring both quadratures (or polarizations) and thereby recording the Q-function of the received states. In this section, we want to demonstrate a key generation system, which is based on postselection of Bob's measurement results. The idea of postselection of continuous variable data was introduced by Hirano et al. \cite{HIR00} and Silberhorn et al. \cite{SIL02b}. The implementation with a discrete set of states was demonstrated in \cite{HIR00,NAM03,HIR03}. The idea for the simultaneous measurement setup was already demonstrated in \cite{LOR04} and further elaborated in~\cite{WEE04}. 

To estimate the efficiency of our experiment of generating key pairs we make
three assumptions. The first concerns the excess noise produced by the quantum
channel. We have shown in the previous section that this excess noise is
always less than 0.02 shot noise units and that we are clearly within the
regime of quantum correlations. We expect that the influence on the key rate
is small for these values of the channel excess noise. Therefore, we neglect
this excess noise in a first approximation and assume a noiseless
quantum channel. A full security analysis, however, will have to take noisy
quantum channels into account.

The second assumption concerns the local oscillator. As already mentioned in
section III we transmit the local oscillator mode and the signal mode through
the quantum channel, and manipulate both by polarization optics. We assume
also in this section that the classical local oscillator mode is not
manipulated by Eve. 

Our third assumption is that Eve performs a collective
attack, which consists of an individual interaction of
Eve's ancilla states with the signals and a coherent measurement onto those
states. Eve is allowed to delay her measurement after the classical
post-processing step in the protocol is completed to optimize her attack. In
this scenario, a lower bound on the secret key rate $G$ is then given by Devetak and Winter \cite{DEV05}
\begin{equation}
  G \geq I_{A:B}-\chi_\mathit{Holevo},
\end{equation}
whereas $I_{A:B}$ denotes the mutual information between Alice and Bob. The
Holevo quantity $\chi_\mathit{Holevo}$ is a function of the states that Eve holds and
quantifies her knowledge about the data. 

With these assumptions, we estimate the secret key rate while using either
direct or reverse reconciliation  combined with postselection. The estimation
of the key rates are based on the derivation given in \cite{HEI05}. In a reverse
reconciliation scheme, the key is built from Bob's measured data to improve
the key rate \cite{GRO03}. This can be achieved by using suitable one-way protocols in the classical post-processing phase of the protocol.

\begin{figure}
\begin{center}
\includegraphics[width=0.35\columnwidth]{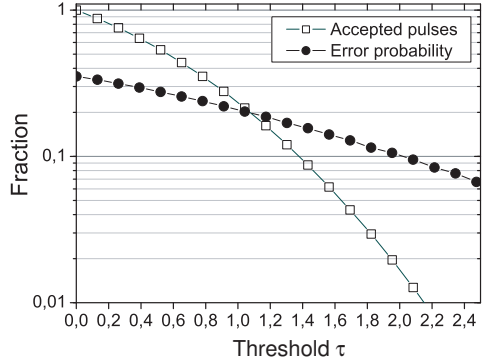}
\end{center}
\caption{Experimental analysis of relative acceptance and average error rate depending on the postselection threshold. Open
  squares: Fraction of accepted states after applying a postselection
  threshold of $\tau$. Black circles: remaining average error rate in the postselected pulses. This measurement was produced with a a channel loss of 51.7\% (cf. Table~\ref{Table_Variances}, last line).}
\label{Figure_ErrorM002}
\end{figure}

The characteristic action of postselection on the data rate can be seen from Fig.~\ref{Figure_ErrorM002}.
The experimental measurement data from the last line in
Table~\ref{Table_Variances} is used to demonstrate the general effect of
postselection on a joint probability distribution $p(A;B)$ derived from our
experiment. After the postselection step, Alice and Bob share correlated data
from which they deduce a binary raw key by assigning a '0' bit value to negative
measurement results, and a '1' bit value to positive measurement results. The x-axis of Fig.~\ref{Figure_ErrorM002} shows the postselection
threshold $\tau$; only data points with $|\mathbf{S_2}|>\tau$ are used to generate the raw key
pair. The open squares show the fraction of the data points which are postselected. The
black circles show the average error rate of the raw key after
postselection. It can be seen that the postselected fraction depends on the
threshold $\tau$ as expected, but also that the average error rate decreases
with increasing threshold, as data points with higher absolute value are less
ambiguous than data points with low absolute value (cf. also \cite{HIR00}). In this sense the plotted
error rate is averaged over all accepted data points, whether the originate
from low absolute values with high error probability or from high
absolute values with low error probability.

A refined version of the postselection procedure uses an analysis which
defines effective binary information channels between Alice and Bob  to estimate the
mutual information $I_{A:B}$ between Alice and Bob and the Holevo quantity
$\chi_\mathit{Holevo}$. It is described in \cite{HEI05} and can be used to determine the
secret key rate $G$ for direct reconciliation using postselection and reverse
reconciliation \cite{GRO02b}. By using these information channels, one can
determine the mutual information and Eve's knowledge about the data separately
for each channel. The secret key rate can be optimized over Alice's input
signal strength. Furthermore, it is possible to include the fact that any
implementation of an error correction scheme cannot reach the theoretical
performance limit given by Shannon \cite{SHA48a, SHA48b}.

Following the calculations in \cite{HEI05}, we can predict the key rates for a
realistic error correction protocol that is assumed to perform as efficient as
the widely used error correction protocol Cascade \cite{BRA94}. The pulse rate for the experiment was 100~kPulse/s, the signal rate was 16.7~kPulse/s due to calibration. For the experiments of Table~\ref{Table_Variances} the key rates are shown in Table~\ref{Table_KeyRates}.
\begin{table}[tbh]
\begin{tabular}{cccc}
State & Quantum  & Key rate  & Key rate\\ 
overlap & channel & DR and & RR and \\
$\langle -\alpha | + \alpha \rangle$ & transmission T & postselection & postselection \\ \hline \hline
0.50 & 45.7\% & 0.0027 & 0.0168 \\ 
0.77 & 45.7\% & 0.0004 & 0.0025 \\ 
0.52 & 48.3\% & 0.0038 & 0.0194 \\ 
0.65 & 48.3\% & 0.0021 & 0.0106 \\ 
0.51 & 65.0\% & 0.0244 & 0.0562 \\ 
\hline \hline
\end{tabular}
\caption{Relative key rates for the experimental data, assuming realistic error correction and no channel excess noise. DR stands for direct reconciliation, RR for reverse reconciliation.}
\label{Table_KeyRates}
\end{table}
With the setup used, clock rates up to 2MPulse/s are feasible, with no need for calibration (vacuum pulses) in the case of key generation, thus much higher secret key rates will be achieved in future experiments.


\section{Conclusions}
We presented an experiment to verify the entanglement intrinsically present in Alice's and Bob's preparation and measurement data in a prepare\&measure quantum key distribution experiment. Under the assumption that the local oscillator cannot be used by Eve to gain any information we built a coherent state measurement setup with high quantum efficiency and low added noise. For two overlapping coherent states prepared by Alice, we show that the joint probability distribution $p(A;B)$ can only be explained by effective entanglement between Alice and Bob. This is the precondition for establishing a secret shared key \cite{CUR04b, CUR05, ACI05}. In addition, by our special measurement setup we reconstruct Husimi's Q-function for the states received by Bob, which is useful in detecting manipulations in the quantum channel. We show that the excess noise of our coherent states is within the measurement accuracy, and less than 2\% of the variance of the shot noise. With this low noise, many attacks by Eve can be ruled out for a large range of transmission losses, allowing longer distance key distribution without compromising the security. By applying postselection to our measurement data, we showed that a secure key can be generated when Eve is restricted to use the beam splitting attack. 


\section{Acknowledgments}
This work was supported by the German engineers society (VDI) and the German
science ministry (BMBF) under FKZ:13N8016, by the network of competence QIP of
the state of Bavaria (A8), the EU-IST network SECOQC and the German Research
Council (DFG) under the Emmy-Noether program. The authors would like to thank
N. Korolkova and D. Elser for valuable discussions.


\bibliographystyle{apsrev}

\end{document}